\newtheorem{zzz}{Lemma}
\newtheorem{eqsetseqn}[zzz]{Lemma}
\newtheorem{conscon}{Theorem}
\newtheorem{simple}[zzz]{Lemma}
\newtheorem{monodromy}[conscon]{Theorem}
\def \for {\ {\rm for}\ }
\def \be {\begin{equation}}
\def \ee {\end{equation}}
\def \ll {\label}
\def \ca {{\cal A}}
\def \ox {\otimes}
\def \unit {{\bf 1}}
\def \complex {{\bf C}}
\documentstyle[12pt,a4]{article}
{\LARGE \bf\author{{\sc  Ladislav   Hlavat\'{y}}
\thanks{Postal   address:
B\v{r}ehov\'{a} 7, 115 19
Prague 1, Czech Republic. E-mail: hlavaty@br.fjfi.cvut.cz}
\\ {\it Department  of  Physics,}
\\ {\it  Faculty  of  Nuclear  Sciences  and
Physical Engineering}}
\title{Algebraic Framework for Quantization of Nonultralocal
Models}

\begin{document}
\bibliographystyle{unsrt}
\maketitle
hep-th/9412142
\abstract{Extension of the braid relations to the multiple braided
tensor product of algebras that can be used for quantization of
nonultralocal models is presented. The Yang--Baxter--type consistency
conditions
as well as conditions for the existence of the
multiple coproduct (monodromy matrix), which can be used for construction of
the
commuting subalgebra, are given. Homogeneous and local algebras
are introduced, and simplification of the Yang--Baxter--type conditions for
them
is shown. The Yang--Baxter--type equations and multiple coproduct conditions
for
 homogeneous and local of order 2
algebras are solved.}
\vskip 2cm
\section {Introduction}
There is a well known procedure for solving a class of nonlinear
diferential equations -- inverse scattering method
\cite{GGKM}. For quantization of integrable field
models like sine--Gordon (SG), nonlinear Schroedinger (NLS), and others, its
quantum counterpart was developed
\nocite{fadskl:doklady78,skl:doklady79,STF:qism,thawil:qism,HWW:qism}
\cite{fadskl:doklady78}--\cite{HWW:qism}.
 The fundamental object of the quantum inverse scattering method is the
quantum transition or monodromy matrix -- the generating function of conserved
quantities.

The  starting point for derivation of the commutation relations of the
elements of the monodromy matrix usually
is the fact that the
Poisson brackets for the Lax matrix $L$ of  classical
models
quantized by the quantum inverse scattering method can be written by virtue of
the classical $r$--matrix as
\be
\{L(x,\lambda)\stackrel{\ox}{,}L(y,\mu)\}
=[r(\lambda,\mu),L(x,\lambda)\ox\unit + \unit\ox L(y,\mu)]
\delta(x-y)\ll{Poisson bracketsL} \ee
where $\{\ \stackrel{\ox}{,}\ \}$ means the Poisson brackets of
elements of $M\times M$ matrices,
$r$ is a matrix $M^2\times M^2$ and $[\ ,\ ]$ means commutator of
matrices. If one denotes the $M^2\times M^2$ matrices
$ L\ox \unit $ and $\unit\ox L$ as $L_1$ and $L_2$
 then an alternative notation (and more appropriate in the
 following) for the set of the Poisson brackets (\ref{Poisson bracketsL}) is
\be
\{L_1(x,\lambda),L_2(y,\mu)\}
=[r_{12}(\lambda,\mu),L_1(x,\lambda) +  L_2(y,\mu)]
\delta(x-y)\ll{PBL} \ee
The  classical transition matrix $\tau(x,y;\lambda)$
is defined by
\be \partial_x \tau(x,y;\lambda) = L(x,\lambda)\tau(x,y;\lambda),
\ \ \ \tau(y,y;\lambda) = \unit
\ll{clasmm} \ee
and its Poisson brackets  read \cite{fad:leshouches}
\be \{\tau(x,y;\lambda)\stackrel{\ox}{,}\tau(x,y;\mu)\}
=[r_{12}(\lambda,\mu),\tau(x,y;\lambda)\ox\tau(x,y;\mu)]
\ee
or
\be \{\tau_1(x,y;\lambda),\tau_2(x,y;\mu)\}
=[r_{12}(\lambda,\mu),\tau_1(x,y;\lambda)\tau_2(x,y;\mu)]
\ee
Quantization of relations (\ref{PBL}) is given by the famous relations
\be R_{12}(\lambda,\mu)L_1^n(\lambda)L_2^n(\mu) =
L_2^n(\mu)L_1^n(\lambda)R_{12}(\lambda,\mu) \ll{rll} \ee
where the matrix $R$ satisfies Yang--Baxter equation and $L^n$ are quantum
analogs of lattice
regularized Lax operators.
The {\em quantum monodromy matrix} is defined in the quantum inverse
scattering method as
\be T^N(\lambda) := L^N(\lambda)
L^{N-1}(\lambda)\ldots L^1(\lambda). \ll{qmm} \ee
and one finds that
the commutation relations of its elements are the same as those for the
elements of
$L^n$.
\be R_{12}(\lambda,\mu)T_1^N(\lambda)T_2^N(\mu) =
T_2^N(\mu)T_1^N (\lambda)R_{12}(\lambda,\mu). \ll{rtt} \ee
In derivation of relations for both classical and quantum monodromy matrices
one
assumes that the models are {\em ultralocal}. For classical
theories it means that the Poisson bracketss of the Lax operators
$L(x,\lambda)$ are proportional to the $\delta-$function but not to
its derivatives. In the quantum theories it means
that the quantized operators $L^n$ sitting in different sites
$n,m$ of the
lattice commute
\be L_1^nL_2^m= L_2^mL_1^n \ \ {\rm for }\  m\neq n. \ll{lnlm} \ee

Many integrable models like e.g. NLS and SG in the space--time
coordinates are ultralocal but there are models like KdV,
non--linear sigma models or SG
in the light--cone coordinates where the Poisson brackets of the $L$--operators
are proportinal to the derivatives of the $\delta$--function.
 Thorough investigation of the classical nonultralocal models can
be found in \cite{mai:npb86}.

The first attempt to deal with quantization of nonultralocal models in the
framework of the quantum inverse scattering method was done in the paper of
Tsyplyaev
\cite{tsy:tmf81}
where a regularization to lattice ultralocal models was chosen and
they were then quantized.

Quantization of relations
\be \{T_1(\lambda), T_2(\mu)\} = [r_{12}(\lambda,\mu),T_1(\lambda)T_2(\mu)]
+T_1(\lambda)s_{12}(\lambda,\mu)T_2(\mu) -
T_2(\mu)s_{12}(\lambda,\mu)T_1(\lambda) \ll{tfrimai} \ee
for classical monodromy matrix of periodic nonultralocal models
that are the consequence of the nonultralocal relations
\[ \{L(x,\lambda)\stackrel{\ox}{,}L(y,\mu)\}
=[r(\lambda,\mu),L(x,\lambda,\mu)\ox\unit + \unit\ox L(y,\mu)]
\delta(x-y)+
\] \[
[s(\lambda,\mu),L(x,\lambda,\mu)\ox\unit - \unit\ox L(y,\mu)]
\delta(x-y)-2s(\lambda,\mu)
\delta'(x-y). \]
was suggested in \cite{frimai}. The quantum version of the relations
\be \{T_1(\lambda)\stackrel{\ox}{,} T_2(\mu)\} = a_{12}T_1(\lambda)T_2(\mu)
 +T_1(\lambda)b_{12}T_2(\mu) -
T_2(\mu)c_{12}T_1(\lambda) -T_1(\lambda)T_2(\mu) d_{12} \ll{abcdclas} \ee
that include relations
(\ref{tfrimai}) is
\be A_{12}(\lambda,\mu)T_1(\lambda)B_{12}(\lambda,\mu)T_2(\mu) =
T_2(\mu)C_{12}(\lambda,\mu)T_1(\lambda)D_{12}(\lambda,\mu) \ll{abcd} \ee

Another step was made in \cite{babbon:qtt} where a classical nonultralocal
model
was discretized in such way that the nonultralocality was preserved and
then it was quantized. The quantum relations for the discretized Lax
operator remain nonultralocal in the sense that  $L^n$ sitting in the
neighbouring sites do not commute.
\be L_1^nL_2^{n+1}= L_2^{n+1}A_{12}L_1^n \ll{lnlnp} \ee

The spectrally dependent version of these relations appeared in the papers
\cite{nijcap:lnp,nijetal:pra92} on
integrable mappings for lattice KdV and related models.
Relations for the quantized Lax operators read
\be R_{12}^+(\lambda,\mu)L_1^n(\lambda)L_2^n(\mu)
 = L_2^n(\mu)L_1^n(\lambda)R_{12}^-(\lambda,\mu)\ll{rlln} \ee
\be L_1^n(\lambda)L_2^{n+1}(\mu)=
L_2^{n+1}(\mu)S_{12}(\lambda,\mu)L_1^n(\lambda) \ll{lnlnp1} \ee
The matrices $R^{\pm},S$ satisfy conditions
\be R_{12}^\pm(\lambda,\mu) R_{13}^\pm(\lambda,\nu) R_{23}^\pm(\mu,\nu)
=R_{23}^\pm(\mu,\nu)
R_{13}^\pm(\lambda,\nu) R_{12}^\pm(\lambda,\mu)
\ll{rrrpm} \ee
\be R_{12}^\pm(\lambda,\mu) S_{13}(\lambda,\nu) S_{23}(\mu,\nu)
=S_{23}(\mu,\nu)
S_{13}(\lambda,\nu) R_{12}^\pm(\lambda,\mu)
\ll{rss} \ee
and
\be R_{12}^-(\lambda,\mu)
S_{12}(\lambda,\mu)=S_{21}(\mu,\lambda)R_{12}^+(\lambda,\mu). \ee
They guarantee that compact relations for the quantum monodromy matrix
(\ref{qmm}) can be
obtained.

The reason why in the ultralocal case the quantum monodromy matrix (\ref{qmm})
satisfies the same relations as
its components $L^n$ is that the algebra defined
by the relations (\ref{rll}) for fixed $n$ admits the matrix coproduct
or, in other words, it can be extended into
the bialgebra. The quantum monodromy matrix then actually is a
representation of the multiple
matrix coproduct of $L$.
Observation made in \cite{hlakun:nulm} is that similar situation occurs
in the examples of quantized nonultralocal models but instead of normal
matrix bialgebras (related to quantum groups)
 we must use their braided analogs (related to quantized braided
groups
\cite{hla:qbg}) because the commutation relations for
neighbouring Lax operators $L^n$ can be interpreted as braiding relations
between copies of an algebra that together with the matrix
coproduct form the braided bialgebra. The quantum monodromy matrix
can again be defined as a representation of the multiple
matrix coproduct of generators.
For this reason we
believe that the multiple braided products of algebras investigated in this
paper  can be used for quantization of nonultralocal models.

\section {Braided matrix bialgebras}\label{bmb}
We shall start with the simple braided product of algebras. Let $\cal A$ is
the complex associative algebra generated by unit element
$\unit_\ca$ and
$L_j^k,\
j,k \in \{1,2,\ldots ,M\}$
satisfying quadratic relations
\be A_{12}L_1B_{12}L_2=L_2C_{12}L_1D_{12} \ll{albl} \ee
where $L=\{L_j^k\}_{j,k = 1}^{M}$ and $A,B,C,D$ are numerical matrices
$M^2\times M^2$ (i.e.
$A_{12}=\{(A)_{i_1i_2}^{j_1j_2}\in \complex \}
_{i_1,i_2,j_1,j_2 = 1}^{M}$
and similarly for $B,\ C,\ D$). These algebras were introduced
in \cite{frimai} and include (algebras of functions on) quantum
groups, quantum supergroups, braided groups, quantized braided
groups, reflection algebras and others.

The consistency conditions for the algebra (\ref{albl}) are \cite{frimai}
\[ [A,A,A]= 0, \hskip 1cm [D,D,D]=0, \]
\be [A,C,C]= 0, \hskip 1cm [D,B,B]=0, \ll{conc}\ee
\[ [A,B^+,B^+]= 0, \hskip 1cm [D,C^+,C^+]=0, \]
\[ [A,C,B^+]= 0, \hskip 1cm [D,B,C^+]=0, \]
where we have introduced (constant) {\em Yang--Baxter commutator}
\be [R,S,T]:=R_{12}S_{13}T_{23}-
T_{23}S_{13}R_{12} \ll{ybc} \ee
and the superscript $X^+$ means $PXP$ where $P$ is the
{\em permutation matrix} $P_{ij}^{kl}=\delta_i^l\delta_j^k$.

The first problem we shall solve is the existence of the
coproduct, i.e. a coassociative map $\Delta : \cal A \rightarrow \cal A \otimes
\cal A $ that is homomorphism of the algebras.
The algebras for which such a map  can be found are called
{\em bialgebras.}

There are two well known classes of algebras of the type (\ref{albl})
where {\em matrix coproduct}
\be \Delta (L_i^j) = L_i^k \otimes L_k^j. \ll{cop} \ee
exists. They are the quantum groups where
$A=D=R,\ B=C=1$ and braided matrix groups \cite{maj:jmp91} where $A=C=R^+,\
B=D=R$. In both cases the consistency conditions (\ref{conc})
reduce to the constant Yang--Baxter quation
\[ [R,R,R]=0\]
for the matrix R.
The important difference between these two classes of bialgebras is
in the definition of the product in $\ca\ox\ca$.

In the quantum groups, the multiplication
in $\cal A \otimes
\cal A $ is defined by the usual way namely

\[ (a  \otimes b)\cdot(c \otimes d) = ac \ox bd \]

i.e.
\[ m_{\cal A \otimes
\cal A }=(m_{\ca} \ox m_{\ca}) \circ (id\ox \tau\ox id), \]
where $\tau$ is the flip operator
\[ \tau(a\ox b)=b\ox a, \]        and $m_{\ca}$ is the
multiplication map in $\ca$.

In the braided groups the multiplication is defined
in a more complicated way, namely
\be m^{\psi}_{\cal A \otimes
\cal A }=(m_{\ca} \ox m_{\ca}) \circ (id\ox \psi\ox id),
\ll{brmul} \ee
where $\psi$ is the so called braiding i.e. isomorphism $\psi :
\ca\ox\cal A \rightarrow \cal A \otimes
\cal A $ satisfying several properties \cite{maj:beyond}. Clearly
$m_{\ca\ox\ca}=m^\tau_{\ca\ox\ca}$.

The algebras for which  homomorphism
$\Delta$ 
into the algebras with the product (\ref{brmul}) can be
introduced are called {braided bialgebras}. The algebras
where it is of the form (\ref{cop}), we shall call {\em braided
matrix bialgebras}.

The question now is, when the algebra (\ref{albl}) is a braided
matrix bialgebra? We shall restrict ourselves to investigating
the braidings of the form \cite{maj:jmp91}
\begin{equation}
\psi    (L_{i}     ^{m}     \otimes     L_{k}     ^{n})    :=
\psi_{ik}^{mn},_{rs}^{jl}(L_j^r \otimes L_l^s)
\label{psirel}
\end{equation}
where
\begin{equation}
\psi_{ik}^{mn},_{rs}^{jl}:=
W_{id}^{aj}Y_{ar}^{lb}Z_{sb}^{cn} \tilde X_{ck}^{md}.
\ll{psi}
\end{equation}
i.e. to the braidings that can be expressed also as the quadratic relation in
$\ca\ox\ca$.
\be
L_1^{(1)}X_{12}L_2^{(2)}=
W_{12}L_2^{(2)}Y_{12}L_1^{(1)}Z_{12}, \ll{lxl}
\ee
where the elements of $L^{(1)},L^{(2)} \in \ca\ox\ca$ are defined as
${L^{(1)}_i}^j=\unit_\ca\ox L_i^j,\ {L^{(2)}i}^j=
L_i^j\ox \unit_\ca,$ and
\[ \tilde X := ((X^{t_2})^{-1})^{t_2} \Longleftrightarrow
X_{il}^{kn}\tilde X_{kj}^{ml}=\tilde
X_{il}^{kn}X_{kj}^{ml}=\delta^m_n\delta^n_j \]
For this type of braiding we get:

\begin{simple} The algebra $\ca$ given by
 (\ref{albl}) can be extended to the braided matrix bialgebra with
the braiding (\ref{lxl})
if
\be C=B^+,\ X=W=Z=B,\ Y^+A=DY. \ll{cbmb} \ee
\label {simple}\end{simple}
Proof: The definition of matrix coproduct (\ref{cop}) can be rewritten
as $\Delta
(L)=L^{(2)}L^{(1)}$ and using (\ref{albl},\ref{lxl},\ref{cbmb}), one can show
that
\be A_{12}\Delta(L_1)B_{12}\Delta(L_2)=
 \Delta(L_2)C_{12}\Delta(L_1)D_{12} \ll{delalbl} \ee
Q.E.D.

The braided bialgebras given by the relations of the form
\be A_{12}L_1B_{12}L_2=
 L_2B_{21}L_1Y_{21}A_{12}Y^{-1}_{12} \ll{gqbg1} \ee
\be
L_1^{(1)}B_{12}L_2^{(2)}=
B_{12}L_2^{(2)}Y_{12}L_1^{(1)}B_{12}, \ll{Gqbg2}
\ee
that follow from (\ref{cbmb}) are a slight generalization of QBG investigated
in
\cite{hla:qbg}. In that case $Y=B^{-1}.$

Extension of this lemma to the multiple braided product will
appear in the next Section.

\section{Multiple braided product of algebras}\label{mpa}

As mentioned in the Introduction,  the crucial object
for construction of ultralocal models
is
the monodromy matrix. It actually is
a representation of the multiple
matrix coproduct $\Delta^N(L)$ of  generators of the quadratic algebra
(\ref{rll})
with trivial braiding (\ref{lnlm}).

The examples in the Introduction suggest that commutation relations for
quantized
Lax operators of nonultralocal models yield algebras with nontrivial
relations between operators localized in different
lattice points i.e. nontrivial braiding,
nevertheless, the monodromy matrix is again given by product of
Lax operators i.e a representation of
the multiple  coproduct of an algebra.
Therefore, it is worth  to investigate the algebras and braidings
that enable construction of multiple coproducts.
In the
following we shall present a generalization of the braided
matrix bialgebras from the previous section.

The algebra $\cal B$ we are going to investigate is generated by
$\unit_{\cal B}$ and
$N\times M^2$ generators
\be (L^{I})_j^k,\ I \in \{1,2,\ldots ,N\}\
j,k \in \{1,2,\ldots ,M\} \ll{lgen}\ee
satisfying quadratic relations
\be
L_1^JX_{12}^{JK}L_2^K=
W_{12}^{JK}L_2^KY_{12}^{JK}L_1^JZ_{12}^{JK}
\ \ \ J,K \in \{1,2,\ldots ,N\}\  \ll{ljxlk} \ee
where $X^{JK},\ W^{JK},\ Y^{JK},\ Z^{JK}$ for fixed
$J,K\in \{1,2,\ldots ,N\}$ are numerical
invertible matrices $M^2\times M^2$ i.e. $X^{JK}=
X_{12}^{JK}=\{(X^{JK})_{i_1i_2}^{j_1j_2}\in \complex\}
_{i_1,i_2,j_1,j_2 = 1}^M$   and similarly for
$ W,\ Y,\ Z$. Throughout this paper {\em no summation over
indices}
$I,J,K,\ldots$ is assumed.

The partition of the set of the generators into $N$ families  of $M\times M$
generators indicates that this algebra can be interpreted as the
{\em multiple braided tensor product} of $N$ algebras ${\cal A}_1\ox
\ca_2\ox \ldots \ox \ca_N$, where the relations (\ref{ljxlk}) for
$J=K$ define the algebras $\ca_J$ and for $J\neq K$ they define the
braid maps $\psi^{JK}: \ca_J \ox \ca_K \rightarrow \ca_K\ox
\ca_J$.
If we want to interpret the algebra (\ref{ljxlk}) as
the multiple braided product of the same algebra $\ca$ given by (\ref{albl})
 then
\be X^{JJ}=B, \ W^{JJ}=A^{-1},\ Y^{JJ}=C,\ Z^{JJ}=D. \ll{mbpcon}
\ee
must hold for all $J$.

The physical importance of the algebra (\ref{lgen},\ref{ljxlk})
consists in possibility to
desribe commutation relations of Lax operators both on finite and
periodical infinite
lattices.

It may happen that the algebra $\cal B$ can be overdetermined by
the relations (\ref{ljxlk}) because  for fixed $J,K,\ J\neq K$ they
actually define two braiding
relations, namely (\ref{ljxlk}) and
\be L_1^JY_{21}^{KJ}L_2^K=
(W_{21}^{KJ})^{-1}L_2^KX_{21}^{KJ}L_1^J(Z_{21}^{KJ})^{-1}
\ll{ljxl} \ee
To guarantee that these two formulae give the
same braidings, we restrict
the structure matrices $X,\ W,\ Y,\ Z$ by the conditions
\be X^{JK}=(Y^{KJ})^+,\ (W^{KJ})^{-1}=(W^{JK})^+,\
(Z^{KJ})^{-1}=(Z^{JK})^+,\ {\rm for}\ J\neq K. \ll{brcon} \ee

The  quadratic algebras defined by relations of the form (\ref{albl}) or
(\ref{ljxlk}) must satisfy
consistency conditions of the Yang--Baxter--type
 that follow from the requirement that no
supplementary higher degree relations are necessary for unique
transpositions of three and more elements \cite{FRT}. For relations of
the form
(\ref{ljxlk})  the conditions read (cf.\cite{frimai})
\be \{ [Z,Z,Z]\} =0,\ \{ [W,W,W]\} =0 \ll {sybe1}\ee
\be \{ [Z,X,X]\} =0,\ \{ [X,X,W]\} =0 \ll {sybe2}\ee
\be \{ [Z,Y^{\ddagger},Y^{\ddagger}]\} =0,\
\{ [Y^{\ddagger},Y^{\ddagger},W]\} =0
\ll {sybe3}\ee
\be \{ [Z,X,Y^{\ddagger}]\} =0,\ \{ [Y^{\ddagger},X,W]\} =0 \ll {sybe4} \ee
where by $\{ [R,S,T]\} =0$ we mean that
\be  [R,S,T]^{J_1J_2J_3} :=
[R^{J_1J_2}S^{J_1J_3}T^{J_2J_3}] =
R_{12}^{J_1J_2}S^{J_1J_3}_{13}T^{J_2J_3}_{23}
-T_{23}^{J_2J_3}S^{J_1J_3}_{13}R^{J_1J_2}_{12}=0
 \ll{sybc} \ee
for $ J_i \in \{1,\ldots,N\}\ \}$, and
\[ (Y^{\ddagger})^{JK}:=(Y^{KJ})^+ = PY^{KJ}P, \]

The equations  (\ref{sybe1}) -- (\ref{sybe4}) can be derived  by transposing
elements of
$L^{J_1},L^{J_2},L^{J_3}$ in
\be
(L_1^{J_1}X^{J_1J_2}_{12}L_2^{J_2})X^{J_1J_3}_{13}X^{J_2J_3}_{23}
L_3^{J_3} =
L_1^{J_1}X^{J_1J_2}_{12}X^{J_1J_3}_{13}(L_2^{J_2}X^{J_2J_3}_{23}
L_3^{J_3}). \ll{l1l2l3} \ee
Comparing expressions obtained by different order of
transpositions and using the properties of the Yang-Baxter commutator
\be [R,S,T]=0\ \Longleftrightarrow [R^{-1},S^{-1},T^{-1}]=0
\ \Longleftrightarrow [T^{\ddagger},R^{\ddagger},S^{\ddagger}]
=0 \ee we get (\ref{sybe1}) - (\ref{sybe4}).

We shall see that for certain type of matrices $X,Y,Z,W$ the
large systems of equations (\ref{sybe1})--(\ref{sybe4}) reduces to much
smaller.
Note e.g. that if $Y^{JK}=(X^{KJ})^+$ then $Y^\ddagger=X$ and the equations
(\ref{sybe3}),
(\ref{sybe4}) are equivalent to (\ref{sybe2}). Another reductions will
appear in the following.

In the algebra (\ref{ljxlk}) the following theorem, that will be
useful for the
construction of the multiple matrix coproduct, holds.
\begin{monodromy} Let the matrices of generators $L^J$ satisfy
(\ref{ljxlk}), where
\be X^{1,K-1}=Z^{1K},\ \ W^{JN}=X^{J+1,N},\ \ll{mcopcn1} \ee
\be Z^{J+1,1}Y^{J1}=1,\ \ Y^{NK}W^{N,K-1}=1, \ll{mcopcn2} \ee
\be Z^{J+1,K}Y^{JK}W^{J,K-1} = X^{J+1,K-1}, \ \ll{mcopcn3} \ee
for $J=1,2,\ldots,N-1,\ \ K=2,3,\ldots,N$.
Then the elements of the matrix
\be T^{N}:=L^NL^{N-1}\ldots L^1 \ll{tn} \ee
satisfy
\be T^{N}_1X^{1N}_{12}T^{N}_2 = W^{NN}_{12}
T^{N}_2Y^{N1}_{12}T^{N}_1Z_{12}^{1,1}. \ll{tnxtn} \ee
\label{monodromy}
\end{monodromy}

{\em Proof:} It consists in essentially direct, even though
sometimes rather involved calculation. We shall present the main
steps of the calculation.

First we shall show that
(the subscript (12) is omitted in the following
formulas)
\be T^{N}_1X^{1N}T^{N}_2 = W^{NN}L^N_2
Y^{NN}T^N_1(2) W^{1,N-1}
L^{N-1}_2T^N_2(2)Y^{11}L^1_1Z^1_1 \ll{step1} \ee
where we introduce
\be T^N_1(K):=\prod_{J=N}^KL_1^JZ^{J,N-K+2}Y^{J-1,N-K+2},
\ll{tau1}\ee
\be T^N_2(K):=\prod_{J=N-K}^1Y^{K-1,J+1}W^{K-1,J}L_2^J,
\ll{tau2}\ee
and
\be \prod_{J=K}^Mx^J:=x^K\ldots x^{M}\ \ee
Indeed, the exchange relations for $L^J$ (\ref{ljxlk}) imply that
if $W^{JN}=X^{J+1,N}$ for $J=1,2,\ldots,N-1$ then
\be T^{N}_1X^{1N}L^N_2 = W^{NN}
L^N_2Y^{NN}T^N_1(2)L^1_1Z^{1N} \ll{hlp1}\ee
and if $ X^{1,K-1}=Z^{1K}$  for $K=2,3,\ldots,N$  then
\be L^1_1Z^{1N}T^{N-1}_2 = W^{1,N-1}
L^{N-1}_2T^N_2(2)Y^{11}L^1_1 Z^{11}\ll{hlp2} \ee
The equation (\ref{step1}) then follows from (\ref{hlp1},\ref{hlp2}).

Due to (\ref{ljxlk}) and (\ref{mcopcn1})--(\ref{mcopcn3}),
the exchange relations between $T^N(K)$ and $L$ read
\[
T^N_1(K) W^{K-1,N-K+1}
L^{N-K+1}_2=\]
\be W^{N,N-K+1}
L^{N-K+1}_2Y^{N,N-K+1}T^N_1(K+1)L_1^{K}Z^{K,N-K+1}\ll{hlpK1}\ee
\be L_1^{K}Z^{K,N-K+1}  T^N_2(K) = W^{K,N-K}
L^{N-K}_2T^N_2(K+1)Y^{K1}L_1^KZ^{K1} \ll{hlpK2} \ee
and thus
\[
T^N_1(K) W^{K-1,N-K+1}
L^{N-K+1}_2T^N_2(K) = \]
\be W^{N,N-K+1}L^{N-K+1}_2Y^{N,N-K+1}
T^N_1(K+1) W^{K,N-K}L^{N-K}_2T^N_2(K+1)Y^{K1}L^1_KZ^{K1}.
\ll{stepK} \ee

The crucial point of the proof is that left--hand side of (\ref{stepK}) is
equal to the the middle
part of the right--hand side up to $K\rightarrow K+1$.
As
\[ T_1^N(N)=L_1^NZ^{N2}Y^{N-1,2},\ \ T_2^N(N)=1, \]
we get from
(\ref{step1}) after $N-1$ steps
\[ T^{N}_1X^{1N}T^{N}_2 =
 (\prod_{K=1}^{N-1} W^{N,N-K+1}L^{N-K+1}_2
Y^{N,N-K+1}W^{N,N-K})\times \]
\be L_1^NZ^{N2}Y^{N-1,2}
W^{N-1,21}L^{1}_2
(\prod_{K=N-1}^1Y^{K,1}L_{1}^KZ^{K,1})
\ll{stepN} = \ee
\[ W^{NN}(\prod_{K=1}^{N-1}L^{N-K+1}_2
Y^{N,N-K+1}W^{N,N-K})L^{1}_2Y^{N1}L^N_1
(\prod_{K=N-1}^1Z^{K+1,1}Y^{K,1}L_{1}^K)Z^1_1 \]
Using (\ref{mcopcn2}) we get (\ref{tnxtn}). Q.E.D.
\vskip 1cm
When the relation (\ref{mbpcon}) holds then the algebra $\cal B$ can be
interpreted  as the
multiple braided tensor product of $N$ algebras ${\cal A}$ and
the conds (\ref{mcopcn1})--(\ref{mcopcn3}) together with $X^{1N}=B,\
Y^{N1}=C$ can be interpreted as conds
for the existence of the multiple coproduct of L. This, on the
other hand, can be used for
construction of the monodromy matrix for nonultralocal models.


A simple way to satisfy the conditions (\ref{sybe1})--(\ref{sybe4}),
(\ref{mcopcn1})--(\ref{mcopcn3}), and (\ref{brcon}) for $W,X,Y,Z$ is to choose
\[ X^{JK}=R(J,K+1),\ Y^{JK}=R(K,J+1)^{+},\]
\be W^{JK}=(R(K+1,J+1)^+)^{-1},
\ \ Z^{JK}=R(J,K),
\ll{xwy} \ee
where $R$ is a "unitary" solution of the Yang--Baxter equation
\[ \{ [R,R,R]\} = 0,\ \ R(K,J)^{-1}=R(J,K)^+\ {\rm for} \ J\neq K. \]
The algebraic relations for $\cal B$ in this case are
a generalization of the braid group relations
\[R_{21}(K+1,J+1)L_1^JR_{12}(J,K+1)L_2^K =
L^K_2R_{21}(K,J+1)L^J_1R_{12}(J,K).\]
Another solution of the conditions (\ref{sybe1})--(\ref{sybe4}),
(\ref{mcopcn1})--(\ref{mcopcn3}), and (\ref{brcon})
is given in the next Chapter.

If the assumptions of the Theorem \ref{monodromy} are strengthened
to \[W^{JK}=X^{J+1,K}=Z^{J+1,K-1},\ Y^{JK}=(X^{J+1,K-1})^{-1} \]
then we can derive the exchange relations also for transitio{n}
matrices
\be T^{L,M}:=L^LL^{L-1}\ldots L^M, \ L>M. \ee
They read
\be T^{L_1M_1}_1X^{M_1,L_2}_{12}T^{L_2M_2}_2=
X_{12}^{L_1+1,L_2}T^{L_2M_2}_2(X_{12}^{L_1+1,M_2-1})^{-1}
T^{L_1M_1}_1X_{12}^{M_1,M_2-1}. \ee
\section{Homogeneous and local algebras}

The structure coefficients in the commutation relations for $L^n,L^m$ in the
examples
of nonultralocal models usually do not depend explicitly both on $n$ and $m$
but only on their difference (the equally distant neighbours commute in
the same way irrespectively of the site). That is why in the following we
shall consider a special case of the algebras from the previous
section, namely such that their structure matrices $X,W,Y,Z$ do not
depend on both $J$ and $K$ but on their difference only
\be X^{JK}=X^{J-K},\ Y^{JK}=Y^{J-K},\ W^{JK}=W^{J-K},\
Z^{JK}=Z^{J-K} \ll{homa} \ee
for $J,K=1,\ldots,N$. We shall call these algebras {\em homogeneous.}

In the homogeneous algebras the
multiple coproduct conditions (\ref{mcopcn1})--(\ref{mcopcn3}) are
equivalent to
\be X^I=Z^{I-1}=W^{I-1} \ {\rm for}\ I=-N+2,\ldots,N-1,\ I\neq1 \ll{xi}
\ee
\be Y^I=(Z^{I+1})^{-1}=(W^{I+1})^{-1} \ {\rm for}\
I=-N+1,\ldots,N-2,\ I\neq -1
\ll{yi} \ee
\be X^1=Z^0Y^{-1}W^0 \ll{x1y1} \ee
and the conds (\ref{brcon}) now read
\be Y^I=(X^{-I})^+,\ \
 Z^{-I}=(Z^I)^\ddagger \Longleftrightarrow (Z^I)^{-1}=(Z^{-I})^+
\for I\neq 0. \ll{zi} \ee
Moreover  we get $Y^0=X^0$ from (\ref{xi},\ref{yi},\ref{zi}).

The reduction of the set of the Yang--Baxter--type equations
for homogeneous algebras describes:
\begin{conscon}
If the structure matrices $X,Y,W,Z$ in (\ref{ljxlk})
satisfy (\ref{homa},\ref{xi},\ref{yi},\ref{zi})   and
$P(X^{1-N})^{-1}P=:Z^N$ then the set of the
Yang--Baxter--type equations (\ref{sybe1}--\ref{sybe4}) is equivalent to
\be [Z^K,Z^{I+K},Z^I]=0,
\ll{homascon1} \ee
\be  [Z^I,Z^I,W^0]=0,\ \ [W^0,Z^K,Z^K]=0,
\ll{homascon2} \ee
\be [W^0,W^0,W^0]=0,\ [X^1,X^1,W^0]=0,\ [Z^0,X^1,X^1]=0,
\ll{homascon3} \ee
\be  [Z^I,Z^I,X^1]=0,\ [X^1,Z^I,Z^I ]=0,\
\ll{homascon4} \ee
where $ 0\leq K \leq N,\
0\leq I\leq N-1,\ 0\leq I+K \leq N $.
\label{conscon}\end{conscon}
For the proof of the theorem we shall use two following lemmas
\begin{eqsetseqn}
If we denote $(R^\#)^{JK} := (R^{\ddagger})^{KJ} = P(R^{KJ})^{-1}P$
and analogously $S^\#,T^\#$, then
(\ref{sybc}) that
\[ [R,S,T]^{IJK}=0
\ \Longleftrightarrow
\ [R^{\#},T,S]^{JIK}=0\ \Longleftrightarrow
\ [S,R,T^{\#}]^{IKJ}=0 \]
\label{eqsetseqn}
\end{eqsetseqn}
{\em Proof:} Follows directly from the definition of Yang--Baxter
commutator (\ref{sybc}).
\\{\em Corollary:}
\[  [R,S,T]^{IJK}=0\Longleftrightarrow [S^{\#},T^{\#},R]^{KIJ}=0\
\Longleftrightarrow\]
\[ \Longleftrightarrow [T,R^{\#},S^{\#}]^{JKI}=0\ \Longleftrightarrow
\ [T^{\#},S^{\#},R^{\#}]^{KJI}=0, \]
\begin{zzz}If  $(Q^\#)^{JK} = Q^{JK}$ then
$ \{[Q,Q,Q]\}=0$ is equivalent to
\[ [Q,Q,Q]^{IJK}=0,\ {\rm for}\  I\geq J\geq K.\]
\label{zzz}\end{zzz}
{\em Proof:}
For any $I',J',K'$ there is a permutation $\pi$ such that
$\pi(I')\geq\pi(J')\geq\pi(K')$, and from Lemma \ref{eqsetseqn}
we get
\[ [Q,Q,Q]^{I'J'K'}=0,\ \Longleftrightarrow
[Q,Q,Q]^{\pi(I')\pi(J')\pi(K')}=0,\]
{\em Proof of the Theorem} \ref{conscon}: First, from
(\ref{xi},\ref{yi},\ref{zi}) we get
\[ Y^{JK}=Y^{J-K}=(X^{K-J})^+ =(X^{KJ})^+ \]
so that $X=Y^{\ddagger}$
and both (\ref{sybe3}) and (\ref{sybe4}) are equivalent to (\ref{sybe2}).

Second, if we
introduce $Z^{-N}=W^{-N}:=X^{1-N}=P(Z^N)^{-1}P$ then due to (\ref{xi})
\[ X^{JK}=Z^{J-K-1}=W^{J-K-1}\ {\rm for }\
J-K\neq 1\]
and  the set of equations (\ref{sybe2}) can be rewritten as
\be [Z,X,X]^{I,J,K} = [Z^{I-J},Z^{I-K-1},Z^{J-K-1}] =
[Z,Z,Z]^{I,J,K+1}=0
\ll{sy21}\ee
for $I,J\neq K+1$,
\be  [X,X,W]^{I,J,K} = [W^{I-J-1},W^{I-K-1},W^{J-K}]=
[W,W,W]^{I-1,J,K}=0 \ee
for $ J,K \neq I-1$, and
\be [Z,X,X]^{K+1,K+1,K} =[Z^0,X^1,X^1] = 0, \ee
\be [X,X,W]^{I,I-1,I-1} = [X^1,X^1,W^0] = 0,\ll{x1x1w0}\ee
\[ [Z,X,X]^{I,J,I-1} = [Z^{I-J},X^1,Z^{J-I}] =0,\]
\[ [Z,X,X]^{J,I,I-1} = [Z^{I-J},Z^{I-J},X^{1}] =0,\]
\[ [X,X,W]^{K+1,J,K} = [Z^{K-J},X^1,Z^{J-K}] =0,\]
\[ [X,X,W]^{K+1,K,J} = [X^1,Z^{J-K},Z^{J-K}] =0,\]
for $I\neq 1,\ K\neq N, I\neq J,\ K\neq J$.
The last four equations are equivalent to
\be [Z^{L},Z^{L},X^1] =0,\ [X^1,Z^{L},Z^{L}] =0,\ \
L=1,2,\ldots,N-1 \ll{zlzlx1}\ee
due to Lemma \ref{eqsetseqn} and $(Z^\#)^{JK}=Z^{JK}$ for $J\neq
K$.

Third, it follows from Lemma \ref{zzz} and $(Z^\#)^{JK}=Z^{JK},\
(W^\#)^{JK}=W^{JK}$
for $J\neq K$ that the set (\ref{sybe1}) is equivalent to
\be [W^{I-J},W^{I-K},W^{J-K}]=0,\ [Z^{I-J},Z^{I-K},Z^{J-K}]=0\ I\geq J\geq K
\ll{sy1}\ee
and this  is equivalent to the system
\be [W^0,W^0,W^0]=0,\
\ [W^0,Z^{I-K},Z^{I-K}]=0,\ [Z^{I-J},Z^{I-J},W^0]=0,\
\ll{wz} \ee
\be [Z^{I-J},Z^{I-K},Z^{J-K}]=0,
\ll{zzzn-1} \ee
for $I\geq J\geq K$, because
$W^J=Z^J$ for $J\neq 0$.

Comparing the sets of equations (\ref{sy21})--(\ref{x1x1w0}),
(\ref{zlzlx1}), (\ref{wz},\ref{zzzn-1}) with
(\ref{homascon1})--(\ref{homascon4}) completes the proof.

In summary, we have found that {\em the homogeneous
algebras admitting the multiple coproduct are determined by  the
matrices}
$W^0,X^1,Z^0,Z^1,\ldots,Z^{N}$ that satisfy
\be  X^1=Z^0(X^1)^+W^0 \ll{zxw} \ee
and  the consistency
conditions (\ref{homascon1})--(\ref{homascon4}).

The relations for the homogeneous  algebra then  read
\be
L_1^J(Z^1_{21})^{-1} L_2^J=
W^0_{12}L_2^J(Z^1_{12})^{-1}L_1^JZ^0_{12}, \ \ 1\leq J\leq N \ll{homa1}
\ee
\be
L_1^{J+1}X^1_{12}L_2^J=
Z^1_{12}L_2^J(Z^2_{12})^{-1}L_1^{J+1}Z^1_{12},\ \ 1\leq J\leq N-1  \ll{homa2}
\ee
\be
L_1^{J+K}Z_{12}^{K-1}L_2^J=
Z^K_{12}L_2^J(Z^{K+1}_{12})^{-1}L_1^{J+K}Z^K_{12},\ \ 1\leq J\leq
N-K,\ 2\leq K\leq N-1,\    \ll{homa3}
\ee

For models with periodic boundary conditions the definition of
homogeneous
algebras must be modified in such way that the relations
(\ref{homa1})--(\ref{homa3}) hold for all $J,K=1,\ldots,N$ and
the summation $J+1,\ J+K$ e.t.c. is done $mod\ N$. Besides that,
the condition of periodicity in the infinite lattices
\[ X^{JK}=X^{J,K+N}=X^{J+N,K},\]
and analogously for $Y,W,Z$, yield in the homogeneous algebras
\be X^{I}=X^{I+N},\ Y^{I}=Y^{I+N},\ Z^{I}=Z^{I+N},\
W^{I}=W^{I+N}.
\ee
{}From (\ref{zi}) and the definition of $Z^N$ we then get
\be Z^{N-I}=P(Z^{I})^{-1}P,\ I=1,\ldots,N-1 \ee
\be Z^{N}=P(X^1)^{-1}P. \ee
\vskip 1cm
Next we want to formalize the fact that
the models we consider are local in the
sense that they have nontrivial commutation relations only for
several neighbouring
operators.
We shall call algebras (\ref{ljxlk}) {\em local of order} $Q$ if
\be L^J_1L_2^K=L_2^KL_1^J \for |J-K| > Q. \ll{locon} \ee
The local algebras of order 0 can  be called {ultralocal} and
they are the ordinary (i.e. unbraided) multiple tensor products of
algebras $\ca$.
Let us stress here that we speak about locality of commutation relations and
not about the locality of hamiltonians.

One can see from (\ref{homa1})--(\ref{homa3}) the homogeneous algebra are
local of order $Q>0$ iff
\[ Z^I =\unit \for I\geq Q \]
and ultralocal iff
\[ Z^I =\unit \for I\geq 1,\ X^1=\unit.  \]

For models with the periodic boundary conditions the definition of local
algebras is modified to the form
\be L^J_1L_2^K=L_2^KL_1^J \for |(J-K) mod N| > Q. \ll{loconper} \ee

\section{Spectrally dependent algebras}
In order that we may apply the above introduced algebras for
quantization of nonultralocal models we must consider spectral dependent
generators and structure matrices i.e. algebras given
by
\be
L_1^J(\lambda_1)X_{12}^{JK}(\lambda_1,\lambda_2)L_2^K(\lambda_2)=
W_{12}^{JK}(\lambda_1,\lambda_2)L_2^K(\lambda_2)
Y_{12}^{JK}(\lambda_1,\lambda_2)L_1^J(\lambda_1)
Z_{12}^{JK}(\lambda_1,\lambda_2)
\ll{ljxlklam} \ee
The importance of these algebra consists in the fact that we can
find a  commuting subalgebra that can be used for construction of
quantum hamiltonian of a model together with conserved
quantities.

It is easy to
check that all formulas in Chapters (\ref{bmb},\ref{mpa}) remain valid when
we replace generators
\[L \rightarrow L(\lambda),\
L_1 \rightarrow L_1(\lambda_1),\ L_2 \rightarrow L_2(\lambda_2)
,\]
matrices
\[ M^{JK}_{12} \rightarrow M^{JK}_{12}(\lambda_1,\lambda_2) \]
for $M=X,Y,W,Z$ and the set of Yang--Baxter commutators (\ref{sybc}) by
\[ \{ [R,S,T]\} := \{
[R,S,T]^{J_1J_2J_3}(\lambda_1,\lambda_2,\lambda_3) :=\]\be
R_{12}^{J_1J_2}(\lambda_1,\lambda_2)S^{J_1J_3}_{13}(\lambda_1,\lambda_3)
T^{J_2J_3}_{23}(\lambda_2,\lambda_3)-
T_{23}^{J_2J_3}(\lambda_2,\lambda_3)S^{J_1J_3}_{13}(\lambda_1,\lambda_3)
R^{J_1J_2}_{12}(\lambda_1,\lambda_2)  \}            \ee
where $J_i \in \{1,\ldots,N\},\ \lambda_i\in \complex $.

Particularly it means that under the "spectrally modified"
conditions of the Theorem 1 the  multiple coproduct of spectrally dependent
generators
\be T^{N}(\lambda):=L^N(\lambda)L^{N-1}(\lambda)\ldots L^1(\lambda)
\ll{tnlam} \ee
satisfies
\be
T^{N}_1(\lambda_1)X^{1N}_{12}(\lambda_1,\lambda_2)
T^{N}_2(\lambda_2) = W^{NN}_{12}(\lambda_1,\lambda_2)
T^{N}_2(\lambda)
Y^{N1}_{12}(\lambda_1,\lambda_2)T^{N}_1(\lambda_1)Z_{12}^{11}
(\lambda_1,\lambda_2). \ll{tnxtnlam} \ee
Beside that  it was shown in \cite{frimai,nijcap:lnp,hla:gafosc} that
if the numerical matrix function $K(\lambda)$ satisfies
\[ (W_{12}^{NN})^{t_1t_2}(\lambda_1,\lambda_2)
K^{t_1}_1(\lambda_1)\tilde X_{12}(\lambda_1,\lambda_2)
K^{t_2}_2(\lambda_2) = \] \be
K^{t_2}_2(\lambda)\tilde Y_{12}(\lambda_1,\lambda_2)K^{t_1}_1(\lambda_1)
((Z_{12}^{1,1}
(\lambda_1,\lambda_2))^{t_1t_2})^{-1}. \ll{refleq}\ee
where $t_1,\ t_2$ mean the transposition in the first
respectively second pair of indices and
\be \tilde X_{12}(\lambda_1,\lambda_2)=
{({((X_{12}^{1N}(\lambda_1,\lambda_2))^{t_1})}^{-1})}^{t_2} ,
\ee
\be \tilde Y_{12}(\lambda_1,\lambda_2)=
{(({(Y_{12}^{N1}(\lambda_1,\lambda_2))^{t_2})}^{-1})}^{t_1},
\ll{btct} \ee
then the algebra elements
$t(\lambda)=K_i^j(\lambda)T_j^i(\lambda)=tr[K(\lambda)T(\lambda)]$
commute
\[ [t(\lambda_1),t(\lambda_2)] = 0. \]

Relations of the spectral--dependent ultralocal algebras
\[ R_{12}(\lambda,\mu)L_1^J(\lambda)L_2^J(\mu)
=L_2^J(\mu)L_1^J(\lambda)R_{12}(\lambda,\mu)
\]
\[ L_1^J(\lambda)L_2^K(\mu)
=L_2^K(\mu)L_1^J(\lambda),\ J\neq K. \]
are special case of (\ref{ljxlklam}) where
\[ X^{JK}=Y^{JK}=\unit,\ W^{IJ}=Z^{IJ}=\unit\ {\rm for}\
I\neq J,\ W^{JJ}=R^{-1},\ Z^{JJ}=R. \]
The quantum nonultralocal model \cite{nijcap:lnp} mentioned in the
Introduction
is an example of the spectrally dependent algebra
that is homogeneous and local of order 1 because
the relations  (\ref{rlln},\ref{lnlnp1}) are special case of
(\ref{homa1})--(\ref{homa3}) where
\[ Z^0=R^-(\lambda_1,\lambda_2),\]
\[ W^0=R^+(\lambda_1,\lambda_2)^{-1},\]
\[ X^1=S(\lambda_1,\lambda_2),\]
\[ Z^I=\unit\  {\rm for} \ 1\leq I\leq N-1.\]

If we want to find a spectrally dependent,
homogeneous algebra that is local of order  $Q> 1$ we must find
matrices $W^0,X^1,Z^0,Z^1,\ldots,Z^{Q-1}$ satisfying the consistency
conditions (\ref{homascon1}--\ref{homascon4}) where $Z^J=\unit$ for
$Q\leq J\leq N$ in the non--periodic, and $Q\leq J\leq N-Q$
in the periodic case.

We shall present a homogeneous algebra that is local of order 2.
For that purpose we need four spectral  dependent matrices
$W^0(\lambda,\mu),Z^0(\lambda,\mu),X^1(\lambda,\mu),
Z^1(\lambda,\mu)$ that satisfy
\be W^0(\lambda,\mu)=PX^1(\mu,\lambda)^{-1}PZ^0(\lambda,\mu)^{-1}
X^1(\lambda,\mu), \ll{wxzx}\ee
and \be  [Z^0,Z^0,Z^0]=0,\ll{z0z0z0}\ee
\be [Z^1,\unit,Z^1]=0, \ll{z1z}\ee
\be [Z^0,Z^1,Z^1]=0,\ [Z^1,Z^1,Z^0]=0, \ll{z0z1z1}\ee
\be [Z^0,X^1,X^1]=0,\
  [X^1,Z^1,Z^1]=0,\ [Z^1,Z^1,X^1]=0,\ \ll{z0x1}\ee
\be [W^0,Z^1,Z^1]=0,\ [Z^1,Z^1,W^0]=0,
\   [X^1,X^1,W^0]=0,\ee
\be [W^0,W^0,W^0]=0, \ll{w0w0w0}\ee
where [ , , ] now means the spectral dependent Yang--Baxter commutator
\[ [R,S,T]
:=R_{12}(\lambda_1,\lambda_2)S_{13}(\lambda_1,\lambda_3)
T_{23}(\lambda_2,\lambda_3)-
T_{23}(\lambda_2,\lambda_3)S_{13}(\lambda_1,\lambda_3)
R_{12}(\lambda_1,\lambda_2).
 \]
The nontrivial commutation
relations of the spectrally dependent,  local of the order 2,
and homogeneous  algebra then  read
\be
L_1^J(\lambda)Z^1_{21}(\mu,\lambda)^{-1} L_2^J(\mu)=
W^0_{12}(\lambda,\mu)L_2^J(\mu)Z^1_{12}(\lambda,\mu)^{-1}
L_1^J(\lambda)Z^0_{12}(\lambda,\mu),
\ll{l2homa1}
\ee
\be
L_1^{J+1}(\lambda)X^1_{12}(\lambda,\mu)L_2^J(\mu)=
Z^1_{12}(\lambda,\mu)L_2^J(\mu)L_1^{J+1}(\lambda)Z^1_{12}(\lambda,\mu),
 \ll{l2homa2}\ee
\be
L_1^{J+2}(\lambda)Z_{12}^1(\lambda,\mu)L_2^J(\mu)=
L_2^J(\mu)L_1^{J+2}(\lambda),
\ll{l2homa3}\ee

To solve the Yang--Baxter--type system of equations
(\ref{z0z0z0}--\ref{w0w0w0}), we
have started with the six--vertex rational solution of the
Yang--Baxter equation
\be Z^0(\lambda,\mu)\! =\! \left( \begin{array}{cccc}
\lambda\! -\! \mu\! +\!\eta & 0 & 0 &0\\
0 &{\lambda\! -\! \mu} & \eta&0\\
0 & \eta & \lambda\! -\! \mu & 0 \\
0 & 0 & 0 & \lambda\! -\! \mu\! +\!\eta
\end{array} \right), \ll{z0} \ee
and diagonal solution of the equation (\ref{z1z})
\be Z^1(\lambda,\mu) = \pmatrix{
1 & 0 & 0 & 0 \cr 0 & b(\lambda) &0  & 0 \cr
0& 0 & c(\mu)& 0 \cr 0 & 0 & 0 & \kappa b(\lambda)c(\mu)\cr } \ll{z1} \ee
where $b,c$ are arbitrary functions and $\kappa,\ \eta$ are
constants. They satisfy the equations (\ref{z0z0z0})--(\ref{z0z1z1})

We have found several matrices $X^1, W^0$ that solve
(\ref{z0x1})--(\ref{w0w0w0}).
The first class of solutions is given by $X^1$ that is equal to
$Z^0$ modified by an arbitrary function $\zeta$
\be X^1(\lambda,\mu) = \pmatrix{
\lambda-\mu+\eta & 0 & 0 & 0 \cr 0 &
\lambda-\mu & \eta\,\zeta(\mu) & 0 \cr
0& \eta & (\lambda-\mu)\zeta(\mu) & 0 \cr 0 & 0 & 0 &
{(\lambda-\mu+\eta)\zeta(\mu)} \cr } \ll{X11} \ee
and from (\ref{wxzx}) we get the twisted  and gauge transformed
rational solution of Yang--Baxter equation
\be W^0(\lambda,\mu)\! =\! \frac{-1}{(\lambda-\mu)^2-\eta^2}
\left( \begin{array}{cccc}
\frac{\lambda-\mu+\eta}{\zeta(\lambda)} & 0 & 0 & 0 \cr 0 &
\frac{\lambda-\mu}{\zeta(\lambda)} &
\eta\frac{\zeta(\mu)}{\zeta(\lambda)} & 0 \cr
0& \eta & (\lambda-\mu)\zeta(\mu) & 0 \cr 0 & 0 & 0 &
{(\lambda-\mu+\eta)}\frac{\zeta(\mu)}{\zeta(\lambda)}
\end{array} \right), \ll{w11} \ee
Note that for $\zeta= 1$ $W^0(\lambda,\mu)=P(Z^0(\mu,\lambda))^{-1}P$.

The second class of solutions is given by
\be X^1(\lambda,\mu) = \pmatrix{
\lambda+\mu+\rho+\eta & 0 & 0 & 0 \cr 0 &
\lambda+\mu+\rho & \eta\sigma\zeta(\mu) & 0 \cr
0& \eta\sigma & \zeta(\mu)(\lambda+\mu+\rho) & 0 \cr 0 & 0 & 0 &
{\zeta(\mu)(\lambda+\mu+\rho + \eta)} \cr } \ll{X12} \ee
where $\sigma^2=1,\ \zeta$ is an arbitrary function and $\rho$ is
a constant and
\be W^0(\lambda,\mu)\! =\! \frac{1}{(\lambda-\mu)^2-\eta^2}
\left( \begin{array}{cccc}
\frac{\lambda-\mu-\eta}{\zeta(\lambda)} & 0 & 0 & 0 \cr 0 &
\frac{\lambda-\mu}{\zeta(\lambda)} &
-\eta\frac{\zeta(\mu)}{\zeta(\lambda)} & 0 \cr
0& -\eta & (\lambda-\mu)\zeta(\mu) & 0 \cr 0 & 0 & 0 &
{(\lambda-\mu-\eta)}\frac{\zeta(\mu)}{\zeta(\lambda)}
\end{array} \right), \ll{w} \ee
For $\zeta= 1$ $W^0(\lambda,\mu)=(Z^0(\lambda,\mu))^{-1}$.

The third class of solutions is
\be X^1(\lambda,\mu) = \pmatrix{
\rho-\mu & 0 & 0 & 0 \cr 0 &
 \rho-\mu & 0 & 0 \cr
0& \eta & \zeta(\mu)(\rho-\mu) & 0 \cr 0 & 0 & 0 &
{\zeta(\mu)(\rho-\mu)} \cr } \ll{X13}\ee
\be W^0(\lambda,\mu)\! =\! \frac{\rho-\mu}
{(\rho-\lambda)[(\lambda-\mu)^2-\eta^2]}
\left( \begin{array}{cccc}
\lambda-\mu-\eta & 0 & 0 &0\\
0 &\frac{(\lambda- \mu)}{\zeta(\lambda)}
& -\eta\frac{\zeta(\mu)}{\zeta(\lambda)} & 0 \cr
0& -\eta & (\lambda-\mu)\zeta(\mu) & 0 \cr 0 & 0 & 0 &
\frac{(\lambda-\mu-\eta)\zeta(\mu)}{\zeta(\lambda)}
\end{array} \right), \ll{w13} \ee
where $\zeta$ is an arbitrary function and $\rho$ is
a constant. Note that this class generalizes the solution given in
\cite{nijcap:lnp}.

For twisted and gauge transformed version of
$Z^0$ one can  obtain an analogical set of $X^1$ and
$W^0$.

The commutation relations for monodromy matrix are
\be
T^{N}_1(\lambda)X^{1}_{12}(\lambda,\mu)
T^{N}_2(\mu) = W^{0}_{12}(\lambda,\mu)
T^{N}_2(\mu)X^{1}_{21}(\lambda,\mu)T^{N}_1(\lambda)
Z_{12}^{0}
(\lambda,\mu).           \ee
for the periodic lattices, and
\be
T^{N}_1(\lambda)T^{N}_2(\mu) = W^{0}_{12}(\lambda,\mu)
T^{N}_2(\mu)T^{N}_1(\lambda)
Z_{12}^{0}(\lambda
,\mu).                      \ee
for the finite lattices.

\section{Conclusions}

Motivated by examples of quantized nonultralocal models
we have extended the concept of the braided bialgebra
to the multiple braided tensor product of algebras and
we have given arguments for its applications
 in the quantizing of nonultralocal models.

Conditions for the existence of
the multiple braided coproduct in the algebra
(\ref{albl}) are given in the Theorem \ref{monodromy}.
The multiple coproducts, when represented by
operators, are quantum monodromy matrices  that  generate
hamiltonian and conserved
quantities of a quantum model.

Further we have introduced the homogeneous and local algebras
because one can expect that the commutation relations of Lax
operators for many
quantized nonultralocal models on lattice depend
only on their distance in the
lattice.
Rather complicated Yang--Baxter--type consistency
conditions for structure matrices
of the braided tensor product of algebras simplify essentially in
this case -- see Theorem \ref{conscon}. Nevertheless, even then
solving the Yang--Baxter--type
set of equations for the structure matrices of the
algebras remain nontrivial task.

We have presented three classes of
solutions of the Yang--Baxter--type equations that
determine a homogeneous algebra that is local of
order 2. They are of the six vertex
form and depend rationally on the spectral parameters.
The commutation relations
for the monodromy matrix are given as well.
\vskip 1cm
{\bf Acknowledgement:} The author is grateful to Anjan Kundu for
introducing him to the problems of nonultralocal models.
\vskip 5mm
{\bf Note:} Shortly after this work was published in the electronic
preprint version (hep--th/9412142), another paper on this topic, motivated by
Chern--Simons theory, appeared
\cite{schw:hep94}. The algebra investigated there
is spectral dependent homogeneous algebra
(\ref{homa1})--(\ref{homa3}) where
$Z^I$ are independent of $I$ for $I\geq 1$.

\end{document}